\documentclass[onecolumn,preprintnumbers,amssymb,article,nofootinbib]{revtex4}
\usepackage[utf8]{inputenc}
\usepackage[english]{babel}
\usepackage[T1]{fontenc}
\usepackage{graphicx,amssymb,amsmath,wasysym}
\usepackage[bookmarks=false]{hyperref}
\usepackage{graphicx}
\usepackage{dcolumn}
\usepackage{bm}

\begin{document}

\begin{center}{\large
    {\bf }}

  \begin{center}
    \textbf{{\Large On the flavor composition of 
        the high-energy neutrinos in IceCube$^*$} }\\

    \vspace{0.3in}

    \textbf{$\textrm{Sergio Palomares-Ruiz
        }^a$~\footnote{(Speaker) Sergio.Palomares.Ruiz@ific.uv.es}, 
      $\textrm{Olga Mena
        }^a$~\footnote{omena@ific.uv.es} and
      $\textrm{Aaron C. Vincent}^{a,
        b}$~\footnote{aaron.vincent@durham.ac.uk \\
    $^*$ {\sl Prepared for the Proceedings of the 37th International
          Conference on High Energy Physics (ICHEP14), Valencia, 
          June 2-9, 2014.} \\ }} 
    \\
    \vspace{0.2in}
    \textsl{${}^a$ Instituto de F\'{\i}sica Corpuscular (IFIC),
      CSIC-Universitat de Val\`encia,\\ Apartado de Correos 22085,
      E-46071 Valencia, Spain}\\
    \vspace{0.1in}
    \textsl{${}^b$ Institute for Particle Physics Phenomenology
      (IPPP), Department of Physics, \\ Durham University,  
      Durham DH1 3LE, United Kingdom } \\
  \end{center}
\end{center}

\begin{abstract}
The IceCube experiment has recently released 3 years of data of the
first ever detected high-energy ($\gtrsim 30$~TeV) neutrinos, which
are consistent with an extraterrestrial origin.  In this talk, we
compute the compatibility of the observed track-to-shower ratio with
possible combinations of neutrino flavors with relative proportion
($\alpha_e:\alpha_\mu:\alpha_\tau$)$_\oplus$.  Although this
observation is naively favored for the canonical ($1:1:1$)$_\oplus$ at
Earth, once we consider the IceCube expectations for the atmospheric
muon and neutrino backgrounds, this flavor combination presents some
tension with data.  We find that, for an astrophysical neutrino
$E_{\nu}^{-2}$ energy spectrum, ($1:1:1$)$_\oplus$ at Earth is
currently disfavored at 92\%~C.L.  We discuss the trend of this result
by comparing the results with the 2-year and 3-year data.  We obtain the
best-fit for ($1:0:0$)$_\oplus$ at Earth, which cannot be achieved
from any flavor ratio at sources with averaged oscillations during
propagation.  Although it is not statistically significant at present,
if confirmed, this result would suggest either a misunderstanding of
the expected background events, or a misidentification of tracks as
showers, or even more compellingly, some exotic physics which deviates
from the standard scenario.
\end{abstract}

\maketitle

\section{Introduction}
\label{sec:introduction}

The first evidence for a high-energy neutrino flux of
extraterrerstrial origin was obtained with a 2-year search in the
IceCube neutrino detector, from May 2010 to May
2012~\cite{Aartsen:2013bka, Aartsen:2013jdh}.  In this period, 28
veto-passing events were recorded (7 tracks and 21 showers) with
deposited energies between $\sim$30~TeV and $\sim$1~PeV.  The
atmospheric neutrino and muon background is expected to be
$10.6^{+5.0}_{-3.6}$ events~\cite{Aartsen:2013jdh}.  This rate, and
also the observed spectrum, is inconsistent with this background
alone, with a significance of $4.1\sigma$.  Recently, an extra year of
data was released, with 2 extra tracks and 7 extra showers in the
$\sim$30~TeV to 2~PeV range, which increases the significance of their
extraterrestrial origin at the $5.7\sigma$
level~\cite{Aartsen:2014gkd}.  The identification of the sources of
this incoming neutrino flux requires disentangling the background from
the signal, with the study of the energy distribution of the observed
events, their correlation with photons and/or protons, their arrival
direction and their flavor composition.  A first detailed discussion
on the flavor composition of these high-energy neutrinos was carried
out in Ref.~\cite{Mena:2014sja}.  In this talk, we highlight the main
results presented there and extend further the discussion with the
3-year results.   

For the 2-year (3-year) data, IceCube expects to see 8.6 (12.1) tracks
from the background~\cite{Aartsen:2013jdh, Aartsen:2014gkd}, whereas
only 7 (9) tracks have been observed above $\sim$30~TeV in deposited
energy.  This would imply that the astrophysical component
overwhelmingly produces showers inside the detector.  However, as a
result of (photo)hadronic interactions, astrophysical neutrinos are
commonly modeled as the decay products of pions, kaons and secondary
muons, and the expectation for the neutrino flavor ratio at the
source\footnote{We use the subscript ``$S$'' to denote the flavor
  composition at the location of the astrophysical sources, before any
  propagation effect takes place, whereas  ``$\oplus$'' represents the
  composition at Earth. In the analysis presented here, we do not
  place any restrictions on the flavor ratios at Earth.} is
$(\alpha_{e, S}:\alpha_{\mu, S}:\alpha_{\tau, S})=$($1:2:0$)$_S$.
However, these neutrinos travel over cosmic distances, so oscillations
are averaged and this ratio becomes $(\alpha_{e, \oplus}:\alpha_{\mu,
  \oplus}:\alpha_{\tau, \oplus}) =$ ($1:1:1$)$_\oplus$ at
Earth~\cite{Learned:1994wg}, which leads to a non-negligible component
of astrophysical tracks. However, the comparison between the expected
background and the observed events indicates that there cannot be a
significant number of astrophysical tracks, so a departure from the
canonical expectation is present (see Ref.~\cite{Chen:2013dza},
however). Deviations of the neutrino flavor ratios from this canonical
expectation have been discussed in the literature, as the default
diagnostic of standard effects (including meson energy losses or muon
polarization~\cite{Rachen:1998fd, Kashti:2005qa, Kachelriess:2006fi,
  Lipari:2007su, Pakvasa:2007dc, Hummer:2010ai}), neutron
decays~\cite{Anchordoqui:2003vc}, deviations from tribimaximal
mixing~\cite{Athar:2000yw, Beacom:2003zg, Serpico:2005bs,
  Lipari:2007su, Pakvasa:2007dc, Esmaili:2009dz, Choubey:2009jq,
  Fu:2012zr, Chatterjee:2013tza, Xu:2014via}, neutrino matter effects
in the source~\cite{Mena:2006eq} and other more exotic
scenarios~\cite{Athar:2000yw, Crocker:2001zs, Beacom:2002vi,
  Barenboim:2003jm, Beacom:2003nh, Beacom:2003eu, Esmaili:2009fk,
  Bhattacharya:2009tx, Bhattacharya:2010xj, Baerwald:2012kc,
  Pakvasa:2012db, Aeikens:2014yga, Illana:2014bda}. 

At the energies under consideration, below a few PeV, there are two
main event topologies in IceCube: muon tracks, associated with a
propagating muon, and electromagnetic or hadronic showers.  Here, we
assess the probability of observing the track-to-shower ratio seen by
the IceCube neutrino telescope as a function of the astrophysical
neutrino flavor composition, and we consider the full range
($\alpha_e:\alpha_\mu:\alpha_\tau$)$_\oplus$ at the detector.
Following Ref.~\cite{Mena:2014sja}, we first outline the calculation
of the muon track and shower event rates in IceCube, after which we
describe our statistical approach.  Then, we present and discuss our
results for the 2-year and 3-year data, summarized in
Figs.~\ref{fig:earth2} and~\ref{fig:earth3}.  We show that, with the
2-year (3-year) data and after accounting for the expected
backgrounds, the canonical combination ($1:1:1$)$_\oplus$ is
disfavored at the 81\% (92\%) confidence level (C.L.) for an
$E_{\nu}^{-2}$ spectrum.  We stress that the new 3-year data follow a
similar proportion of tracks and showers, so the observation does not
seem to be an statistical fluctuation, as shown in
Fig.~\ref{fig:CLtime}.

\section{Neutrino events in IceCube}
\label{sec:events}

At these energies, the IceCube events consist of two type of event
topologies: muon tracks and showers.  In both cases, we consider the
deposited energy to be equal to the sum of the energies of all the
showers in the event and whenever a muon is produced, we neglect the
small amount of energy deposited along the muon track. We also
neglect the small suppression of the light yield in hadronic showers
due to the presence of more neutral particles~\cite{Wiebusch}.  We do
not take into account the error in the determination of the deposited
energy and we use the effective detector masses for each flavor and
type of interaction as a function of the neutrino energy, instead of
the detector mass as a function of the deposited
energy\footnote{Whereas the effective masses as a function of the
  neutrino energy were published in Ref.~\cite{Aartsen:2013jdh}, the
  effective mass as a function of the deposited energy is not publicly
  available.}, which is the quantity which is actually measured.
Although these approximations are not appropriate when performing a
spectral analysis, they introduce very small errors in an analysis
with a single and wide energy bin like the one we consider here.
Overall, we have checked that all these approximations have little
impact on our results.

Showers are induced by both $\nu_e$ and $\nu_\tau$ charge current (CC)
interactions, as well as by neutral current (NC) interactions of
neutrinos of all three flavors.  The total number of showers (sh)
produced by NC interactions for any neutrino (and analogously
antineutrino) flavor $i$ reads
\begin{equation}
\label{eq:nc}
N^{\rm sh,NC}_{\nu_i} = T \, N_A \,
\int^{\infty}_{E_{\rm min}} dE_\nu \, M^{\rm NC} (E_\nu) \,
Att_{\nu_i}(E_\nu) \, \frac{d\phi_{\nu_i}(E_\nu)}{dE_\nu} \times
\int^{y_{\rm max}}_{y_{\rm min}} dy \, \frac{d\sigma^{\rm NC} (E_\nu,y)}{dy} ~,
\end{equation}
where $E_\nu y = (E_\nu-E'_\nu)$ is the shower energy and $E'_\nu$ is
the energy of the outgoing neutrino, with $y_{\rm min} = E_{\rm
  min}/E_\nu$ and $y_{\rm max} = {\rm min} \{ 1, E_{\rm max}/E_\nu
\}$.  The minimum (maximum) deposited energy in this analysis is
$E_{\rm min} = 30$~TeV ($E_{\rm max} = 2$~PeV).  The differential NC
cross section is ${d\sigma^{\rm NC}}/{dy}$, $T$ is the observation time,
$M^{\textrm{NC}}$ is the energy-dependent effective detector mass for 
NC interactions, $N_A=6.022 \times 10^{23} {\rm g}^{-1}$,
$Att_{\nu_i}$ is the attenuation/regeneration factor due to the
absorption and regeneration of $\nu_i$ when traversing the Earth and
$d\phi_{\nu_i}/dE_\nu$ is the $\nu_i$ flux. 

Using the same notation, the total number of CC $\nu_e$ (and
analogously $\bar\nu_e$) induced showers can be written as   
\begin{equation}
\label{eq:cce}
N^{\rm sh, CC}_{\nu_e} = T \, N_A \, \int^{\infty}_{E_{\rm min}}
dE_\nu \, M^{\rm CC}_{\nu_e} (E_\nu) \, Att_{\nu_e}(E_\nu) \, 
\frac{d\phi_{\nu_e}(E_\nu)}{dE_\nu} \times \int^1_0 dy \,
\frac{d\sigma^{\rm CC}_{\nu_e} (E_\nu,y)}{dy} \times
\Theta\left(E_{\rm max}-E_\nu\right) ~. 
\end{equation}

For $\nu_\tau$ (and analogously for $\bar\nu_\tau$), the total number
of shower events induced by CC interactions with an hadronic tau decay
mode is given by~\cite{Dutta:2000jv} 
\begin{eqnarray}
\label{eq:cctau_had} 
N^{\rm sh,CC-had}_{\nu_\tau} & = & T \, N_A \, \int^{\infty}_{E_{\rm
    min}} dE_\nu \,  M^{\rm CC}_{\nu_\tau} (E_\nu) \, Att_{\nu_\tau}(E_\nu) \,
\frac{d\phi_{\nu_\tau}(E_\nu)}{dE_\nu} \int^1_0 dy \,
\frac{d\sigma^{\rm CC}_{\nu_\tau} (E_\nu,y)}{dy} \int^1_0 dz \,
\frac{dn(\tau \rightarrow {\rm had})}{dz}  \nonumber\\
& & \times \Theta\left(E_\nu(y+(1-y)(1-z)) - E_{\rm
  min}\right) \times \Theta\left(E_{\rm max} -
E_\nu(y+(1-y)(1-z))\right) ~,   
\end{eqnarray}
where the total hadronic shower energy is the sum of the hadronic
energy from the broken nucleon, $E_\nu y$, and the hadronic energy
from the tau decay, $E_\nu (1-y) (1-z)$, where $z=
E^\prime_\nu/E_\tau$, with $E^\prime_\nu$ the energy of the neutrino
from the decay.  The spectrum of the daughter neutrino in hadronic
$\tau$ decays is $dn(\tau \rightarrow {\rm had})/dz$. 

The number of showers produced by the electronic decay of the tau
lepton, $N^{\rm sh,CC-em}$, is written in an analogous
way~\cite{Dutta:2000jv},  
\begin{eqnarray}
\label{eq:cctau_em} 
N^{\rm sh,CC-em}_{\nu_\tau} & = & T \, N_A \, \int^{\infty}_{E_{\rm
    min}} dE_\nu \, M^{\rm CC}_{\nu_\tau} (E_\nu) \, Att_{\nu_\tau}(E_\nu) \,
\frac{d\phi_{\nu_\tau}(E_\nu)}{dE_\nu} \int^1_0 dy \,
\frac{d\sigma^{\rm CC}_{\nu_\tau} (E_\nu,y)}{dy} \int^1_0 dz_e \,
\frac{dn(\tau \rightarrow {\rm e})}{dz_e}  \\ 
& & \times \Theta \left(E_\nu(y+(1-y)z_e\right) - E_{\rm min})
\times \Theta \left(E_{\rm max} -E_\nu(y+(1-y)z_e\right)) ~, \nonumber  
\end{eqnarray}
where the electron distribution from tau decays is $dn(\tau
\rightarrow e)/dz_e$, with  $z_e= E_e/E_\tau$, and $E_e$ is the
electron energy.  The total number of showers produced by $\nu_\tau$
CC interactions (and equivalently by $\bar\nu_\tau$), $N^{\rm
  sh,CC}_{\nu_\tau}$, is the sum of the purely hadronic and
hadronic/electromagnetic showers.      

Tracks are induced by muons from $\nu_\mu$ and $\nu_\tau$ CC
interactions.  The energy deposited in the detector comes dominantly  
from the hadronic shower, so we consider the muon track as a tag for
this type of events.  Thus, the total number of contained-vertex
track-like (tr) events from $\nu_\mu$ (and analogously from
$\bar\nu_\mu$) is 
\begin{equation}
\label{eq:tracksmu}
N^{\rm tr}_{\nu_\mu} = T \, N_A \, \int^{\infty}_{E_{\rm min}} dE_\nu
\, M^{\rm CC}_{\nu_\mu} (E_\nu) \, Att_{\nu_\mu}(E_\nu) \,
\frac{d\phi_{\nu_\mu}(E_\nu)}{dE_\nu} \int^{y_{\rm max}}_{y_{\rm min}}
dy \, \frac{d\sigma^{\rm CC}_{\nu_\mu} (E_\nu,y)}{dy} ~.   
\end{equation}

The total number of muon tracks produced by CC $\nu_\tau$ (and
$\bar\nu_\tau$) interactions, $N^{\rm tr}_{\nu_\tau}$, followed by tau
decays ($\tau \to \nu_\tau \nu_\mu\mu$), is given by
\begin{equation}
\label{eq:trackstau}
N^{\rm tr}_{\nu_\tau} = T \, N_A \, \int^{\infty}_{E_{\rm min}} dE_\nu
\, M^{\rm CC}_{\nu_\tau} (E_\nu) \, Att_{\nu_\tau}(E_\nu) \,  
\frac{d\phi_{\nu_\tau}(E_\nu)}{dE_\nu} \int^{y_{\rm max}}_{y_{\rm
    min}} dy \, \frac{d\sigma^{\rm CC}_{\nu_\tau} (E_\nu,y)}{dy} \,
Br(\tau \rightarrow \mu) ~,    
\end{equation}
where $Br(\tau \rightarrow \mu)$ is the branching ratio of tau decays
into muons.

For the neutrino and antineutrino differential cross sections we use
the \texttt{nusigma} neutrino-nucleon scattering MonteCarlo
code~\cite{Blennow:2007tw}, which uses the CTEQ6 parton distribution
functions~\cite{Pumplin:2002vw, Pumplin:2005rh}.  We use the IceCube
effective masses $M^{\rm CC}_{\nu_i}$ and $M^{\rm
  NC}$~\cite{Aartsen:2013jdh}.  The attenuation/regeneration factors
have been computed for each flavor and for neutrinos and antineutrinos
independently following Refs.~\cite{Naumov:1998sf, Iyer:1999wu,
  Rakshit:2006yi}.  We have not included the small
correction~\cite{Dutta:2002zc} due to the secondary $\nu_\mu$ and
$\nu_e$ flux produced by $\nu_\tau$ interactions~\cite{Beacom:2001xn}.
The attenuation/regeneration factor in the above equations is the
average factor for the whole sky, and thus it only depends on the
incoming neutrino energy.  We assume the astrophysical neutrino flux
to be given by the same power law and the same normalizations,
$E_{\nu}^{-\gamma}$, for the three neutrino and antineutrino flavors.
Throughout this letter, we consider $\gamma = 2$ as our default value,
which is in good agreement with the data~\cite{Aartsen:2013jdh,
  Aartsen:2014gkd}.

\section{Statistical analysis}
\label{sec:analysis}

The fractions of electron, muon and tau neutrinos produced in 
astrophysical sources are denoted as $\{\alpha_{i, S}\}$.  After
propagation, averaged neutrino oscillations cause the flavor ratio at
Earth to be $\{\alpha_{j,\oplus}\} = \sum_{k, i} |U_{jk}|^2 \, |U_{ik}|^2
\{\alpha_{i, S}\}$, where $U$ is the neutrino mixing matrix for which
we use the latest $\nu$\textit{fit}
results~\cite{GonzalezGarcia:2012sz} (see also
Refs.~\cite{Tortola:2012te, Fogli:2012ua}).  For $\{\alpha_{i, S}\} =$
($1:2:0$)$_S$, this yields a flavor ratio at Earth of
($1.04:0.99:0.97$)$_\oplus$, very close to the tribimaximal
expectation, ($1:1:1$)$_\oplus$.

For a given combination $\{\alpha_{i,\oplus}\}$, the total number of
events produced by astrophysical neutrinos is  
\begin{equation}
\label{eq:Na}
N_{\rm a}(\{\alpha_{i,\oplus}\}) = \alpha_{e, \oplus} \, (
N_{\nu_e}^{\rm sh,CC} + N_{\nu_e}^{\rm sh,NC} ) + \alpha_{\mu, \oplus}
\, (N^{\rm tr}_{\nu_\mu} + N^{\rm sh,NC}_{\nu_\mu}) + \alpha_{\tau,
  \oplus} \, (N^{\rm tr}_{\nu_\tau} + N^{\rm sh,CC}_{\nu_\tau} +
N^{\rm sh,NC}_{\nu_\tau}) ~,
\end{equation}
where we implicitly assume the sum of neutrino and antineutrino
events.  The proportion of muon tracks\footnote{We have checked that
  the fraction of tracks and showers predicted by the IceCube
  collaboration for different astrophysical
  spectra~\cite{Aartsen:2014gkd} agrees with our expectations for the
  ($1:1:1$)$_\oplus$ flavor ratio.} is  
\begin{equation}
p_{\rm a}^{\rm tr}(\{\alpha_{i, \oplus}\}) = \frac{1}{N_{\rm
    a}(\{\alpha_{i,\oplus}\})} \left( \alpha_{\mu,\oplus} \,
N^{\rm tr}_{\nu_\mu} + \alpha_{\tau,\oplus} \, N^{\rm tr}_{\nu_\tau}
\right) ~,   
\label{ptdef}
\end{equation}
and conversely for showers, $p_{\rm a}^{\rm sh}
(\{\alpha_{i,\oplus}\}) \equiv 1 - p_{\rm a}^{\rm tr}
(\{\alpha_{i,\oplus}\})$.  

For the 2-year (3-year) data, the IceCube collaboration estimated the
background of atmospheric muons and neutrinos to be $b_\mu = 6 \pm
3.4$ ($b_\mu = 8.4 \pm 4.2$) and $b_\nu = 4.6^{+3.7}_{-1.2}$ ($b_\nu =
6.6^{+5.9}_{-1.6}$), respectively~\cite{Aartsen:2013jdh,
  Aartsen:2014gkd}.  In the results presented below, we take the
background events to be Poisson-distributed, but we do not include the
quoted systematic errors.  We note that even if we consider the lower
end of the $1\sigma$ intervals, only about two tracks would be allowed
to be of astrophysical origin, in both the 2-year and the 3-year data
samples.  Let us notice that, even in that case, the number of expected
astrophysical showers would be much larger than that of tracks, a
factor of about 10 larger.  We have checked that this does not change
the best-fit value, but it just slightly reduces the significance of
our results.  Additionally, neutrinos from atmospheric charmed meson
decays could represent a few extra background events.  Given the
uncertainty in this prediction (see, e.g., Ref.~\cite{Enberg:2008te}),
we consider this case separately and use a benchmark
component~\cite{Enberg:2008te}.  For the fraction of background
showers and tracks in the $30~{\rm TeV}-2$~PeV energy range, we use
the numbers quoted by the IceCube collaboration: tracks account for
$69\%$ of the conventional atmospheric neutrino event rate, $19\%$ of
the prompt atmospheric neutrino event rate and $90\%$ of the events
induced by atmospheric muons~\cite{Aartsen:2014gkd}.  We have also
checked that the uncertainties in the ratio of tracks to showers from
atmospheric neutrinos, as computed with different initial fluxes, do
not change our results in a significant way.  For instance, using the
high-energy atmospheric neutrino fluxes of
Refs.~\cite{Sinegovsky:2011ab, Petrova:2012qf, Sinegovskaya:2013wgm},
the fraction of tracks induced by the conventional flux is $\sim
50\%$.  This would only weaken our conclusions by changing the
C.L. contours by a few percent.

\begin{figure}[t]
\includegraphics[width=.85\textwidth]{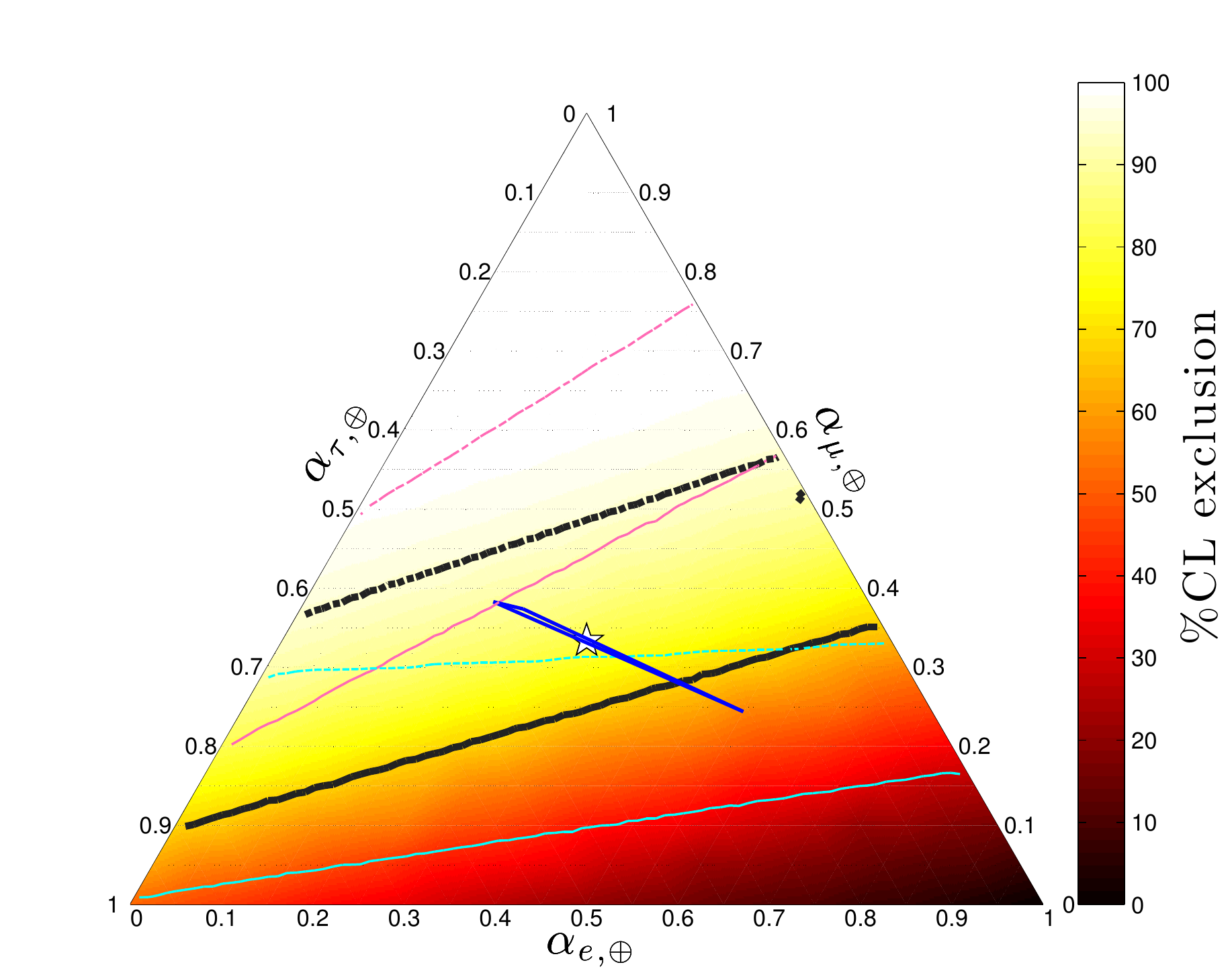}
\caption{\sl \textbf{\textit{Ternary plot of the exclusion
      C.L. for all possible flavor combinations ($\alpha_{e, \oplus}:
        \alpha_{\mu,\oplus}: alpha_{\tau,\oplus}$) as seen at Earth,
        given the 7 tracks and 21 showers observed at IceCube after 2
        years.}}  The lower right corner 
  corresponds to 100\% electron neutrinos, the upper corner is 100\%
  muon neutrinos, and the lower left corner to 100\% tau neutrinos.
  The central sliver in blue corresponds to the possible flavor
  combinations for astrophysical neutrinos, after oscillations have
  been averaged during propagation.  The best-fit is the darkest point,
  ($1:0:0$)$_\oplus$.  The white star corresponds to
  ($1:1:1$)$_{\oplus}$, which is expected from a ($1:2:0$)$_S$
  combination at the source.  The color scale indicates the exclusion 
  C.L. given an $E_{\nu}^{-2}$ spectrum of incoming neutrinos.  Solid
  (dashed) lines show 68\%~C.L. (95\%~C.L.) contours, cyan for
  $E_{\nu}^{-1}$, thick black for $E_{\nu}^{-2}$ and pink for
  $E_{\nu}^{-3}$ spectra. From Ref.~\cite{Mena:2014sja}.}   
\label{fig:earth2}
\end{figure}

\begin{figure}[t]
\includegraphics[width=.85\textwidth]{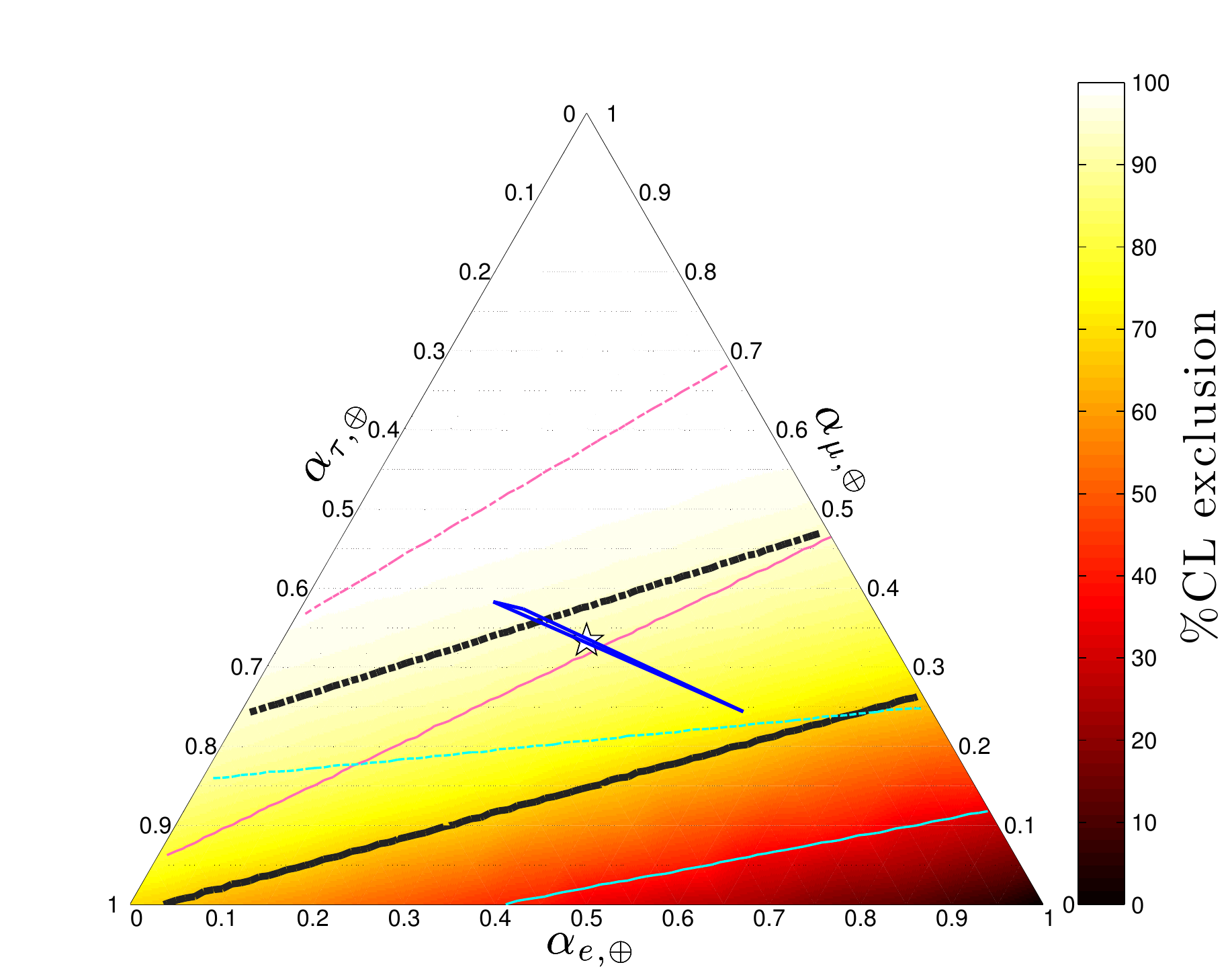}
\caption{\sl \textbf{\textit{Same as Fig.~\ref{fig:earth2}, but given
      the 9 tracks and 28 showers observed at IceCube after 3 years.}}
  The best-fit is still ($1:0:0$)$_\oplus$, and the canonical flavor
  ratio ($1:1:1$)$_{\oplus}$ is disfavored at a slightly higher C.L.} 
\label{fig:earth3}
\end{figure}

The likelihood of observing $N_{\rm tr}$ tracks and $N_{\rm sh}$
showers, for a given combination $\{\alpha_{i,\oplus}\}$ and a total 
number of astrophysical neutrinos $N_{\rm a}$, is 
\begin{eqnarray}  
\mathcal{L}(\{\alpha_{i,\oplus}\}, N_{\rm a}| N_{\rm tr}, N_{\rm sh})
& = & e^{- (p_{\rm a}^{\rm tr} N_{\rm a} + p_{\mu}^{\rm tr} b_\mu +
  p_{\nu}^{\rm tr} b_\nu)} \, \frac{(p_{\rm a}^{\rm tr} \, N_{\rm a} +
  p_{\mu}^{\rm tr} \, b_\mu + p_{\nu}^{\rm tr} \, b_\nu)^{N_{\rm
      tr}}}{N_{\rm tr}  !} \nonumber \\
& & \times \, e^{- (p_{\rm a}^{\rm sh} N_{\rm
    a} + p_{\mu}^{\rm sh} b_\mu + p_{\nu}^{\rm sh} b_\nu)} \,\frac{(
  p_{\rm a}^{\rm sh} \, N_{\rm a} + p_{\mu}^{\rm sh} \, b_\mu +
  p_{\nu}^{\rm sh} \, b_\nu)^{N_{\rm sh}}}{N_{\rm sh} !} ~,      
\end{eqnarray}
where $p_{\nu}^{\rm tr} = 0.69$ ($p_{\nu}^{\rm sh} = 1 - p_{\nu}^{\rm
  tr}$) is the fraction of tracks (showers) in the atmospheric
neutrino background and $p_{\mu}^{\rm tr} = 0.9$ ($p_{\mu}^{\rm sh} =
1 - p_{\mu}^{\rm tr}$) is the fraction of tracks (showers) in the
atmospheric muon background~\cite{Aartsen:2014gkd}.  Since the total
number of events produced by astrophysical neutrinos is not of
interest in this analysis, $N_{\rm a}$ can be treated as a nuisance
parameter and can be set to the value $N_{\rm a}^{\rm  max}
(\{\alpha_{i,\oplus}\})$ which maximizes
$\mathcal{L}(\{\alpha_{i,\oplus}\}, N_{\rm a} | N_{\rm tr}, N_{\rm
  sh})$ for $\{\alpha_{i,\oplus}\}$, yielding  $\mathcal{L}_{\rm
  p}(\{\alpha_{i,\oplus}\} | N_{\rm tr}, N_{\rm sh}) \equiv
\mathcal{L}(\{\alpha_{i,\oplus}\}, N_{\rm a}^{\rm max}
(\{\alpha_{i,\oplus}\}) | N_{\rm tr}, N_{\rm sh})$.

We define the log-likelihood ratio
\begin{equation}
\lambda(N_{\rm tr},N_{\rm sh} | \{\alpha_{i,\oplus}\}) = -2 \ln
\left( \frac{\mathcal{L}_{\rm p}(\{\alpha_{i,\oplus}\} | N_{\rm tr},
  N_{\rm sh})}{\mathcal{L}_{\rm p}(\{\alpha_{i,\oplus}\}_{\rm max} |
  N_{\rm tr},N_{\rm sh})} \right) ~,  
\end{equation}
where $\{\alpha_{i,\oplus}\}_{\rm max} $ is the combination of
neutrino flavors that maximizes the likelihood of observing $N_{\rm
  tr}$ tracks and $N_{\rm sh}$ showers.  The p-value for a given
combination $\{\alpha_{i,\oplus}\}$ is   
\begin{equation}
p(\{\alpha_{i,\oplus}\}) = \sum_{N_{\rm tr},N_{\rm sh}} P(N_{\rm tr},
N_{\rm sh} | \{\alpha_{i,\oplus}\}) ~,   
\label{eq:pvalue}
\end{equation}
where $P(N_{\rm tr}, N_{\rm sh} | \{\alpha_{i,\oplus}\}) \equiv
\mathcal{L}_{\rm p}(\{\alpha_{i,\oplus}\} | N_{\rm tr}, N_{\rm sh})$
is the probability of observing $N_{\rm tr}$ tracks and $N_{\rm sh}$
showers given the flavor ratio $\{\alpha_{i,\oplus}\}$ and $N_{\rm
  a}^{\rm max} (\{\alpha_{i,\oplus}\})$, and the sum runs over all
combinations of $N_{\rm tr}$ and $N_{\rm sh}$ which satisfy
$\lambda(N_{\rm tr}, N_{\rm sh} | \{\alpha_{i,\oplus}\}) >
\lambda(N_{\rm tr} = 7, N_{\rm sh} =  21 | \{\alpha_{i,\oplus}\})$.
Although we use the exact definition of the p-value,
Eq.~(\ref{eq:pvalue}), we note that the test statistic $\lambda$
asymptotically approaches a $\chi^2$ distribution with two degrees of
freedom (see the bottom panel in Fig.~\ref{fig:CLtime}).  The p-value
can easily be translated into an exclusion C.L.:
$C.L. (\{\alpha_{i,\oplus}\}) = 1 - p (\{\alpha_{i,\oplus}\})$.

\section{Results}
\label{sec:results}

\begin{figure}[t]
\includegraphics[width=.8\textwidth]{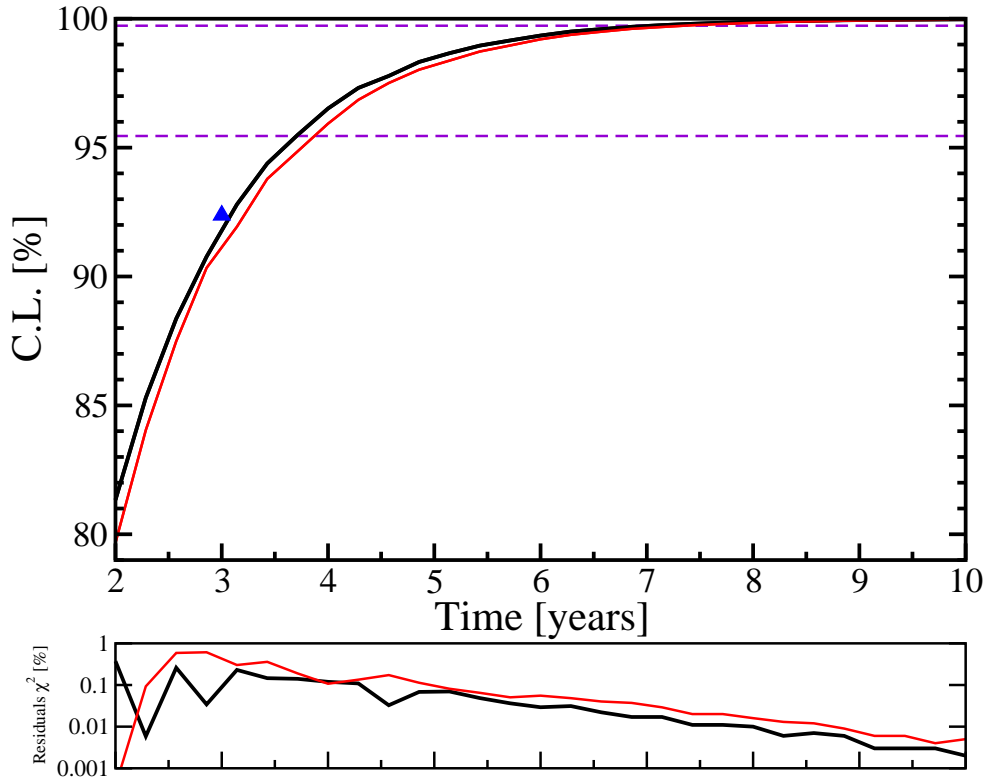}
\caption{\sl \textbf{\textit{Top panel: Evolution of the C.L. of
      exclusion of the canonical flavor ratio ($1:1:1$)$_\oplus$ for
      an $E_{\nu}^{-2}$ spectrum, computed by using the 2-year data
      and the background expectation and extrapolating them in time}}.
  The black (red) line represents the case without the (with the
  benchmark) prompt atmospheric neutrino backgrond.  The blue triangle
  indicates the actual result with the 3-year data.  The two dashed
  lines represent the $2\sigma$ and $3\sigma$ C.L.
  \textbf{\textit{Bottom panel: Residuals with respect to a $\chi^2$
      distribution with two degrees of freedom}}, showing that indeed
  the actual distribution of the test statistic $\lambda$ is very
  close to a $\chi^2$ distribution.}    
\label{fig:CLtime}
\end{figure}

Using Eq.~(\ref{eq:pvalue}), we compute the exclusion limits for all
combinations of $\{\alpha_{i,\oplus}\}$, without any restrictions on
the flavor ratios at Earth.  We show the results for the 2-year data
in Fig.~\ref{fig:earth2} and in Tab.~\ref{tab:cltabEarth} we
quantitatively state how disfavored the canonical flavor ratio
($1:1:1$)$_\oplus$ is.  The color scale shows the exclusion
C.L. assuming an $E_{\nu}^{-2}$ astrophysical spectrum with the same
normalization for all three flavors, which describes well the data in
the $30~{\rm TeV}-2$~PeV energy range~\cite{Aartsen:2013jdh}.  Lines
show the 68\% and 95\%~C.L. limits, which we illustrate for three
different spectra.  For the 2-year data, the ($1:1:1$)$_\oplus$
scenario is disfavored at 81\%~C.L. for an $E_{\nu}^{-2}$ spectrum.
Harder spectra are more constrained, since a larger flux of
$\nu_\mu$'s and $\nu_\tau$'s at high energies leads to the production
of more muons.  We note that the best-fit point is ($1:0:0$)$_\oplus$,
which cannot be obtained from any flavor ratio at sources assuming
averaged oscillations during propagation.   

Beyond the conventional $\pi/K$ atmospheric neutrino background, the 
effect of an atmospheric charm component is also shown in
Tab.~\ref{tab:cltabEarth}, where we see that the changes are not
important.

The results for the 3-year data are depicted in
Fig.~\ref{fig:earth3}, where we use the same convention for colors and
lines as in Fig.~\ref{fig:earth2}.  In this case, the
($1:1:1$)$_\oplus$ scenario is disfavored at 92\%~C.L. for an
$E_{\nu}^{-2}$ spectrum.  Aside from not being yet statistically very 
significant, we note that the 3-year data follows the trend we would
expect if the 2-year results were not a statistical fluctuation.  This
is quantitatively shown in Fig.~\ref{fig:CLtime}, where we compute the
evolution of the C.L. for ($1:1:1$)$_\oplus$ by scaling up in time the
astrophysical and background events of the 2-year data (with and
without the prompt atmospheric background).  The blue triangle
represents the result with the 3-year data without including the
prompt atmospheric background, which is very close to the extrapolation
from the 2-year data.

\begin{table}[t]
\begin{tabular*}{0.5\linewidth}{@{\extracolsep{\fill}}c | c c c} 
$d \phi_\nu /d E_\nu \propto$ & $E_\nu ^{-1}$ & $E_\nu ^{-2}$ & $E_\nu
  ^{-3} $  \\ \hline 
$\pi/K$            & 96\% & 81\% & 52\% \\
$\pi/K$ + charm    & 95\% & 80\% & 53\% \\ \hline
$\pi/K$ (3-yr data)& 99\% & 92\% & 70\% 
\end{tabular*}
\caption{\sl \textbf{\textit{C.L. limits for the ($1:1:1$)$_{\oplus}$
      flavor ratio observed at Earth.}}  The three columns represent
  three possible assumptions for the index of the power-law energy
  spectrum of the astrophysical neutrinos.  ``$\pi/K$'' includes the
  conventional atmospheric muon and neutrino background and ``$\pi/K$
  + charm'' additionally includes the benchmark flux of ``prompt''
  neutrinos from the decay of charmed mesons in the atmosphere.  The
  two upper rows refer to the 2-year data~\cite{Aartsen:2013jdh} and
  the last one to the 3-year data~\cite{Aartsen:2014gkd}.}
\label{tab:cltabEarth}
\end{table}

\section{Summary and discussion}
\label{sec:discussion}

Although the statistical power of the high-energy events seen at
IceCube remains low, the fact that the observed number of tracks is
smaller than the expectation from the atmospheric muon and neutrino
backgrounds allows us to place moderate constraints on the flavor
ratios of the astrophysical neutrinos.  If these are assumed 
to have an unbroken $E_{\nu}^{-2}$ energy spectrum and they are
allowed to have any flavor combination, the ($1:1:1$)$_\oplus$ ratio
at Earth is disfavored at 81\%~C.L (92\%~C.L.) with the 2-year
(3-year) data.  For other spectra, the limits are presented in
Tab.~\ref{tab:cltabEarth}.

Let us also note that for the best-fit for the power-law index of the
astrophysical spectrum quoted by IceCube using the events above
60~TeV (deposited energy), $E_{\nu}^{-2.3}$~\cite{Aartsen:2014gkd}, 
($1:1:1$)$_\oplus$ at Earth is disfavored at 86\%~C.L. with the 3-year
data.  It is compelling to note that significant limits are
potentially at hand.  Indeed, in the 3-year
data~\cite{Aartsen:2014gkd}, the proportion of tracks and showers
is similar to that in the 2-year data.  If the ratio of 1 track per 3
showers above 30~TeV holds for future observations, ($1:1:1$)$_\oplus$
could be disfavored at $3\sigma$~C.L. for an $E_{\nu}^{-2}$ spectrum
after a total of 8 years, as shown in Fig.~\ref{fig:CLtime}.  If this
trend continues, we are faced with several potential implications: (a)
the atmospheric background has been overestimated; (b) some tracks have
been misidentified as showers; (c) the main mechanism of astrophysical
neutrino production is \textit{not} purely hadronic interactions; (d)
no flavor combination at the source provides a good fit to the data
and hence, the observed flavor ratios are due to some non-standard
physics which favors a dominant $\nu_e + \bar{\nu}_e$ composition at
Earth, for instance as in some scenarios of neutrino
decay~\cite{Crocker:2001zs, Beacom:2002vi, Baerwald:2012kc,
  Pakvasa:2012db}, {\sl CPT} violation~\cite{Barenboim:2003jm},
pseudo-Dirac neutrinos~\cite{Beacom:2003eu, Esmaili:2009fk,
  Pakvasa:2012db} or sterile neutrino altered dispersion relations due
to shortcuts in an extra dimension~\cite{Aeikens:2014yga}; or (e)
the neutrino cross sections are different from the standard
expectation at high energies, as in some models of TeV
gravity~\cite{Illana:2014bda}. 

The first very high-energy events detected by IceCube have opened the
door to the era of neutrino astronomy.  Even with such a small sample,
the event topology could provide compelling information on the
production, propagation and detection of neutrinos at high energies.
Nevertheless, these searches are statistically limited.  Therefore, a
future high-energy extension of the IceCube detector and the planned
KM3NeT telescope~\cite{KM3NeT} could be crucial in order to have the
potential to firmly establish the origin and composition of these
neutrinos.

\section*{Acknowledgments}
\label{sec:acknowledgments}
We thank Claudio Kopper for clarifying discussions about the IceCube
detector and data. SPR is supported by a Ram\'on y Cajal
contract and by the Spanish MINECO under grant FPA2011-23596 and by
GVPROMETEOII/2014/049.  OM is supported by the Consolider Ingenio
project CSD2007--00060, by PROMETEO/2009/116, by the Spanish Grant
FPA2011--29678 of the MINECO.  ACV was supported by FQRNT and European
contract FP7-PEOPLE-2011-ITN.  The authors are also partially
supported by PITN-GA-2011-289442-INVISIBLES.  SPR is also partially
supported by the Portuguese FCT through the projects
PTDC/FIS-NUC/0548/2012 and CFTP-FCT Unit 777
(PEst-OE/FIS/UI0777/2013), which are partially funded through POCTI
(FEDER).

\newpage

\bibliographystyle{unsrt}
\bibliography{flavorsICHEP14}

\end{document}